\documentclass[11pt,a4paper]{article}
\pdfoutput=1

\usepackage{jcappub}
\usepackage[utf8]{inputenc}
\usepackage{url}
\usepackage{natbib}

\title{Reconstructing a $f(R)$ theory from the $\alpha$-Attractors}

\author[a]{T.~Miranda,}
\author[a,b]{J.~C.~Fabris,}
\author[a]{and O.~F.~Piattella,}

\affiliation[a]{Department of Physics, Universidade Federal do Esp\'irito Santo,\\
avenida Fernando Ferrari 514, 29075-910 Vit\'oria, Esp\'irito Santo, Brazil}
\affiliation[b]{National Research Nuclear University “MEPhI”,\\ Kashirskoe sh. 31, Moscow 115409, Russia}

\emailAdd{tays.andrade@aluno.ufes.br}
\emailAdd{oliver.piattella@pq.cnpq.br}
\emailAdd{fabris@pq.cnpq.br}

\abstract{
We show an analogy at high curvature between a $f(R) = R + aR^{n - 1} + bR^2$ theory and the $\alpha$-Attractors. We calculate the expressions of the parameters $a$, $b$ and $n$ as functions of $\alpha$ and the predictions of the model $f(R) = R + aR^{n - 1} + bR^2$ on the scalar spectral index $n_{\rm s}$ and the tensor-to-scalar ratio $r$. We find that the power law correction $R^{n - 1}$ allows for a production of gravitational waves enhanced with respect to the one in the Starobinsky model, while maintaining a viable prediction on $n_{\rm s}$. We numerically reconstruct the full $\alpha$-Attractors class of models testing the goodness of our high-energy approximation $f(R) = R + aR^{n - 1} + bR^2$. Moreover, we also investigate the case of a single power law $f(R) = \gamma R^{2 - \delta}$ theory, with $\gamma$ and $\delta$ free parameters. We calculate analytically the predictions of this model on the scalar spectral index $n_{\rm s}$ and the tensor-to-scalar ratio $r$ and the values of $\delta$ which are allowed from the current observational results. We find that $-0.015 < \delta < 0.016$, confirming once again the excellent agreement between the Starobinsky model and observation.}


\arxivnumber{1707.06457}

\begin{document}
\maketitle

\flushbottom


\section{Introduction}

After almost 40 years, the inflationary theory \cite{Starobinsky:1980te, Guth:1980zm, Linde:1981mu, Linde:1983gd} has become an important piece of our understanding of the primordial universe, given its ability to solve some of the puzzles plaguing the standard cosmological model (e.g. the flatness and the horizon problems) and given its ability to make predictions on the primordial power spectrum, which can be tested through observation of the cosmic background radiation (CMB) \cite{Hinshaw:2012aka, Ade:2015lrj}. Since the first inflationary solution, proposed by Starobinsky in Ref.~\cite{Starobinsky:1980te}, see also \cite{Starobinsky:1981zc, Starobinsky:1982ee}, many other proposals have been put forward.\footnote{In Ref.~\cite{Martin:2013nzq} 193 single-field inflationary models are analysed.} These have been classified and statistically analysed in many works, see e.g. Refs.~\cite{Bezrukov:2011gp, Martin:2013nzq, Martin:2013tda, Vennin:2015vfa, Vennin:2015egh, Martin:2016ckm}, and the result is that data favour the simplest category of inflationary models: single-field slow-roll inflation, with a plateau potential. The typical representative of this class of models is the Starobinsky model \cite{Starobinsky:1980te}, which can be expressed within the framework of $f(R)$ theories \cite{Sotiriou:2008rp} as the following quadratic correction to the Einstein-Hilbert action:
\begin{equation}\label{Staromodel}
	f(R) = R + \frac{R^2}{6M^2}\;,
\end{equation}
where $M$ is an arbitrary energy scale, typically $M \sim 10^{13}$ GeV \cite{Liddle:1993ch}. As any $f(R)$, the model \eqref{Staromodel} also can be transformed in a scalar-tensor theory in the Jordan frame and, through a conformal transformation, to a minimally coupled (to the conformally transformed metric) canonical scalar field. 

The predictions of the Starobinsky model on the fundamental inflationary parameters, i.e. the scalar spectral index $n_{\rm s}$ and the tensor-to-scalar ratio $r$ are:
\begin{equation}\label{nsrconstr}
n_{\rm s} - 1 = -\frac{2}{N}\;, \qquad r = \frac{12}{N^2}\;,
\end{equation}
at the leading order of powers of $1/N$, where $N$ is the number of e-folds and whose typical value is $N \gtrsim 60$, which is required in order to solve the flatness and the horizon problems \cite{weinberg2008cosmology}. Choosing $N = 60$ in Eq.~\eqref{nsrconstr} we get 
\begin{equation}\label{nsrconstrStaroN60}
n_{\rm s} = 0.967\;, \qquad r = 3.33\times 10^{-3}\;.
\end{equation} 
These predictions are in very good agreement with the latest Planck results \cite{Ade:2015lrj}:
\begin{equation}
	n_{\rm s} = 0.968 \pm 0.006\;, \qquad r < 0.11\;,
\end{equation}
the former constraint being at 68\% confidence level (CL) and the latter at 95\% CL. The looseness of the second constraint is due to our current inability to detect the B-mode of CMB polarisation, which carries crucial information about the primordial gravitational wave background and thus, through its spectrum, on the energy scale of Inflation. 

The looseness of the constraint on $r$ might be threatening to the primacy of the Starobinsky model in the panorama of all the inflationary, single-field, models of Inflation for if, for example, a sufficiently large $r$ was eventually detected then the Starobinsky model would be ruled out. In this sense it is interesting to explore new theoretical scenarios which do not pin the value of $r$ alongside with that of $n_{\rm s}$, as it happens for the Starobinsky model, cf. Eq.~\eqref{nsrconstr}.

As an example of these scenarios, recently a new class of inflationary models dubbed $\alpha$-Attractors has been proposed in Refs.~\cite{Kallosh:2013hoa, Galante:2014ifa, Kallosh:2015lwa}. These provide the following predictions on the inflationary parameters:
\begin{equation}\label{alphaAttpred}
n_{\rm s} = 1 - \frac{2}{N}\;, \qquad r = \frac{12\alpha}{N^2}\;,
\end{equation} 
where $\alpha > 0$ and, as in Eq.~\eqref{nsrconstr}, the leading order only in the expansion in powers of $1/N$ has been retained. Remarkably, the predictions in the above Eq.~\eqref{alphaAttpred} are very similar to those in Eq.~\eqref{nsrconstr} of the Starobinsky model, so that the excellent agreement with observation as for the parameter $n_{\rm s}$ is maintained while being there freedom to tune the gravitational wave production through the parameter $\alpha$. When the latter is unity, one recovers the Starobinsky model, whereas when $\alpha \to \infty$ one recovers the inflationary paradigm of a non-interacting, massive canonical scalar field \cite{Kallosh:2013hoa}.

The $\alpha$-Attractors theories are most naturally formulated in the context of Supergravity with logarithmic K\"ahler potential. In this sense, a comparison with the Starobinsky model would be more direct and clearer if we had an $f(R)$ theory which could provide the same predictions as in Eq.~\eqref{alphaAttpred}. In particular, we address the following question: how should we modify the Starobinsky model in the $f(R)$ framework in order to have a larger production of gravitational waves? Therefore, we reconstruct a $f(R)$ theory which could mimic the $\alpha$-Attractors scalar potential, given in Eq.~\eqref{alphattpotential}, in the following way. From the latter we obtain a differential equation for $f(R)$, cf. Eq.~\eqref{fequationbeta}, already found in Ref.~\cite{Odintsov:2016vzz}, and which provides the corresponding $f(R)$ theory. Unfortunately, this equation can be solved analytically in order to obtain an explicit form of $f(R)$ only for very specific values of $\alpha$, as done in Ref.~\cite{Odintsov:2016vzz}. Therefore, we analyse its asymptotic behaviour for high energies and show that a solution of the type $f(R) = R + aR^{n-1} + bR^2$, i.e. a power law extension of the Starobinsky model, does exist. Motivated by this asymptotic solution, we study in more detail the model $f(R) = R + aR^{n-1} + bR^2$ calculating its predictions on the inflationary parameters $n_{\rm s}$ and $r$. We find that the parameters $a$, $b$ and $n$ can all be related to one single parameter $\beta$ and that the model is viable for $-0.02 < \beta < 0.8$.  The parameter $\beta$ is related to the $\alpha$ of the $\alpha$-Attractors by $\beta = 1 - 1/\sqrt{\alpha}$. In the appendix we address again Eq.~\eqref{fequationbeta} solving it numerically and showing the goodness of our asymptotic approximation.

Moreover, motivated by Ref.~\cite{Motohashi:2014tra} we also study in more detail the model $f(R) = \gamma R^{2 - \delta}$, calculating analytically its predictions on the inflationary parameters $n_{\rm s}$ and $r$. We find that the $f(R) = \gamma R^{2 - \delta}$ models are viable for $-0.015 < \delta < 0.016$ and do not share with the $\alpha$-Attractors the ability of moving vertically in the plane $n_{\rm s}$ \textit{vs} $r$. 

It must be said that the above-proposed corrections to the Einstein-Hilbert action are valid only for Inflation, thus on very large energy scales, since it has been shown in \cite{Clifton:2005aj} that corrections of the type $R^{1 + \epsilon}$ require $0 \le \epsilon \le 7.2 \times 10^{-19}$, from the Mercury perihelion precession observations. 

The paper is organised as follows. In Sec.~\ref{sec:Staromodel} we briefly emphasise the importance of $f(R)$ theories, and especially of the Starobinsky model, in the context of Inflation via a simplified dynamical system analysis, and we introduce the $\alpha$-Attractors class of models. Then, from the $\alpha$-Attractors potential we derive the differential equation for $f(R)$, cf. Eq.~\eqref{fequationbeta}, which allows to put in correspondence a $f(R)$ theory with the $\alpha$-Attractors. In Sec.~\ref{sec:extStaroalpha} we analyse in detail the model $f(R) = R + aR^{n-1} + bR^2$ and show its predictions on the inflationary parameters. In Sec.~\ref{Sec:singpowlaw} we address in detail the model $f(R) = \gamma R^{2 - \delta}$ and show the analytic derivation of its prediction on the inflationary parameters. Finally, in Sec.~\ref{sec:concl} we present our conclusions. In the Appendix~\ref{App} we numerically analyse in detail Eq.~\eqref{fequationbeta}.

Throughout the paper we use $G = c = 1$ units.
 
 
\section{$f(R)$ theory, Starobinsky model and Inflation}\label{sec:Staromodel}

We start this section by briefly reviewing how an $f(R)$ theory is able to provide a viable inflationary model. Let's start with the following action:
\begin{equation}\label{fRaction}
	S = \frac{M_{\rm Pl}^2}{2}\int d^4x\sqrt{-g}f(R)\;,
\end{equation}
where $M_{\rm Pl}$ is the Planck mass. Following e.g. Ref.~\cite{Barrow:1988xh}, see also Ref.~\cite{Tsujikawa:2014rta}, one can show that the above $f(R)$ action \eqref{fRaction} is equivalent to the Einstein-Hilbert one with a minimally coupled (to the conformally transformed metric) canonical scalar field $\chi$, defined as follows:
\begin{equation}\label{chidef}
\chi = \sqrt{\frac{3}{2}}M_{\rm Pl}\ln\left(\frac{df}{dR}\right)\;, 
\end{equation}
and subject to the following potential:
\begin{equation}\label{Upot}
U = \frac{M_{\rm Pl}^2}{2(df/dR)^2}\left(R\frac{df}{dR} - f\right)\;.
\end{equation}
At this point one can already see why the Starobinsky model \cite{Starobinsky:1979ty} $f(R) = R + R^2/(6M^2)$ is special. Assume for example a $R + \gamma R^n$ theory, with $n > 0$ and where $\gamma$ is some energy scale at which the correction becomes relevant. At high energy scales $\gamma R^n \gg R$, where we expect Inflation to take place, the potential $U$ becomes:
\begin{equation}\label{UpotRn}
U = \frac{M_{\rm Pl}^2(n-1)}{2\gamma n^2}R^{2-n}\;.
\end{equation}
The Starobinsky case $n = 2$ then provides a plateau, which is the ideal situation for a slow-roll inflationary phase to take place. If $n > 2$ the potential initially increases and then goes to zero asymptotically, which might be bad from the point of view of Inflation because the slow-rolling scalar field would need to overcome a potential barrier. For $n < 2$ the potential grows unbound, but still there is the possibility for it to satisfy the slow-roll conditions. See Refs. \cite{Maeda:1988ab, Barrow:1991hg, Broy:2014xwa}.

\subsection{Dynamical system perspective}

Let us analyse the above-described asymptotic behaviours from a dynamical system perspective. Assume a spatially flat Friedmann-Lema\^itre-Robertson-Walker (FLRW) metric: 
\begin{equation}\label{FLRWmetric}
	ds^2 = -dt^2 + a(t)^2\delta_{ij}dx^idx^j\;.
\end{equation}
It is not difficult to cast the evolution equations of $f(R)$ gravity in absence of matter as the following dynamical system:
\begin{equation}\label{RdotHdot}
	\begin{cases}
		6Hf''\dot{R} = Rf' - f - 6f'H^2\\
        6\dot{H} = R - 12H^2
	\end{cases}\;,
\end{equation}
where $H \equiv \dot{a}/a$, the dot denotes derivation with respect to the cosmic time and the prime denotes derivation with respect to $R$. Note that the second equation of system \eqref{RdotHdot} is the very definition of $R$ for metric \eqref{FLRWmetric} and we have assumed $f''$ and $H$ different from zero. 

In order to have an inflationary phase for large $R$ (larger than a certain scale $M^2$), we need $\dot{H} \approx 0$. From the second equation of system~\eqref{RdotHdot} this implies $R \approx 12H^2$ and thus $\dot{R} \approx 0$. This happens if, from the first equation of system~\eqref{RdotHdot}, one has:
\begin{equation}
	\frac{f'}{f} \approx \frac{2}{R}\;, \qquad \mbox{or} \qquad \frac{Rf'/2 - f}{6Hf''} \sim 0\;.
\end{equation}
The first condition leads us to
\begin{equation}
	f(R) \sim R^2\;,
\end{equation}
which is the Starobinsky $R^2$ correction to the Einstein-Hilbert action. The second condition tells us that $6Hf''$ must grow more rapidly than $Rf'/2 - f$ for large $R$. This can be achieved via a $R^n$ correction with $n > 2$, but then the problem that we have mentioned earlier appears: the scalar field has to climb a potential barrier in order for Inflation to end. Recently, an investigation of the inflationary dynamics generated by an inflaton field climbing up a potential has been investigated in Refs. \cite{Jinno:2017jxc, Jinno:2017lun}, where the authors present viable scenarios. Another recent interesting proposal is Ref.~\cite{Amin:2015lnh} where a logarithmic correction to the Einstein-Hilbert action is investigated. In this case, the divergence is linear since the $R$ term dominates but the presence of the logarithmic correction produces a plateau at intermediate energies where Inflation might take place.

The potential \eqref{Upot} for the Starobinsky model can be computed exactly and has the following form:
\begin{equation}\label{Staropotential}
U(\chi) = \frac{3}{4}M^2M_{\rm Pl}^2\left(1 - e^{-\sqrt{2/3}\chi/M_{\rm Pl}}\right)^2\;.
\end{equation}
This potential presents a plateau for high values of the inflation field, as shown in Figs.~\ref{Fig:Uet}.

\subsection{The $\alpha$-Attractors: E-models}

The subclass of $\alpha$-Attractors called E-Models \cite{Kallosh:2015lwa} is described by the following scalar field potential:
\begin{equation}\label{alphattpotential}
U_E(\chi) = \frac{3}{4}M^2M_{\rm Pl}^2\alpha\left(1 - e^{-\sqrt{2/(3\alpha)}\chi/M_{\rm Pl}}\right)^2\;.
\end{equation}
Clearly, the Starobinsky potential \eqref{Staropotential} is recovered for $\alpha = 1$. In Fig.~\ref{Fig:Uet} we display the behavior of the above potential for different choices of $\alpha$. 

Applying the definitions given in Eqs.~\eqref{chidef} and \eqref{Upot} to the E-Models potential of Eq.~\eqref{alphattpotential}, it is not difficult to obtain the following differential equation for $f(R)$:
\begin{equation}\label{fequationbeta}
Rf' - f = \frac{3M^2}{2(1-\beta)^{2}}\left(f' - f^{'\beta}\right)^2\;,
\end{equation}
where we have defined
\begin{equation}
\beta \equiv 1 - \frac{1}{\sqrt{\alpha}}\;.
\end{equation}
Note that, since $\alpha > 0$, then $\beta < 1$. When $\alpha = 1$, i.e. $\beta = 0$, one can easily check that the Starobinsky model is solution of Eq.~\eqref{fequationbeta}. When $\alpha \to \infty$, i.e. $\beta \to 1$, it is easy to see that from Eq.~\eqref{alphattpotential} we get a $\chi^2$ potential and the corresponding $f(R)$ theory can be found by solving the equation:
\begin{equation}
Rf' - f = \frac{3}{2}M^2f'^2\ln^2(f')\;.
\end{equation}
Finally, note that $f = R$ turns Eq.~\eqref{fequationbeta} into a identity. Therefore, Eq.~\eqref{fequationbeta} can be seen as an equation which determines the correction to the usual Einstein-Hilbert action which is able to reproduce the inflationary dynamics of the $\alpha$-Attractors E-models. 

Unfortunately, Eq.~\eqref{fequationbeta} cannot be solved analytically for a generic $\alpha$ and we analyse it numerically in the appendix. On the other hand, if there exists a $f(R)$ theory reconstructing the same inflationary dynamics as the $\alpha$-Attractors, then at high energies the former must behave as the Starobinsky model, i.e. $f(R) \propto R^2$, because the $\alpha$-Attractors potential displays a plateau at high energies and, in the framework of $f(R)$ theories, this is realized only by a $f(R) \propto R^2$ correction.

Let us show this explicitly, assuming a $f(R) = \gamma R^n$ theory, where $\gamma >0$ is an arbitrary energy scale and $n > 1$, and substituting it into Eq.~\eqref{fequationbeta}:
\begin{equation}\label{gammaRneq}
\frac{2(n - 1)(1 - \beta)^2}{3M^2\gamma n^2}R^{2 - n} = \left[1 - (\gamma n)^{\beta - 1}R^{(\beta - 1)(n - 1)}\right]^2\;.
\end{equation}

Since $\beta < 1$ and $n > 1$, the exponent $(\beta - 1)(n - 1)$ is negative. Therefore, in the limit $\gamma R^{n - 1} \to \infty$, we get from Eq.~\eqref{gammaRneq}:
\begin{equation}\label{gammaRneq2}
\frac{2(n - 1)(1 - \beta)^2}{3M^2\gamma n^2}R^{2 - n} \to 1\;.
\end{equation}
In order to satisfy this limit, $n$ must be equal to $2$. This result confirms that the asymptotic behavior of a $f(R)$ theory which aims to reproduce the $\alpha$-Attractors dynamics must go as $R^2$ for large $R$. We then propose this ansatz:
\begin{equation}\label{polcorr}
	f(R) = R + aR^n + bR^2\;,
\end{equation} 
where $a > 0$ and $b > 0$ are arbitrary energy scales and $n > 1$. Let's substitute Eq.~\eqref{polcorr} into Eq.~\eqref{fequationbeta}:
\begin{eqnarray}\label{fequationnewmodel}
	\frac{a(n - 1)R^n + bR^2}{(1 + anR^{n - 1} + 2bR)^2} = \frac{3M^2}{2(1 - \beta)^2}\left[1 - (1 + anR^{n - 1} + 2bR)^{\beta - 1}\right]^2\;,
\end{eqnarray}
and consider the high energy limits:
\begin{equation}
	aR^{n - 1} \gg 1\;, \quad  \mbox{and} \quad bR \gg 1\;.
\end{equation}
If $n < 2$, the dominant contributions in Eq.~\eqref{fequationnewmodel} are those proportional to $R^2$ and we get from them:
\begin{equation}
	b = \frac{(1 - \beta)^2}{6M^2}\;.
\end{equation}
For $\beta = 0$ we then recover the correct form of the coefficient in Eq.~\eqref{Staromodel} for the Starobinsky model. One can check that in Eq.~\eqref{fequationnewmodel} the next-to-leading power in $n$, being $n < 2$, is $n$ itself, whereas the next-to-leading power in $\beta$, being $\beta < 1$, is $\beta + 1$. Therefore, we can approximate Eq.~\eqref{fequationnewmodel} as follows:
\begin{eqnarray}
	a(n - 1)R^n + bR^2 \sim \frac{3M^2}{2(1 - \beta)^2}\left[4b^2R^2 + a^2n^2R^{2(n - 1)} + 4banR^n - 2(2bR)^{\beta + 1}\right]\;.
\end{eqnarray}
We equate the sub-dominant powers and their respective coefficients, obtaining:
\begin{equation}
	n = \beta + 1\; \qquad a = (2b)^\beta\;.
\end{equation}
When $\beta = 0$, the $aR^{n}$ contribution of Eq.~\eqref{polcorr} is thus reincorporated into the usual Einstein-Hilbert term of the action and we are left with the Starobinsky model. The comparison between powers that we have just done in Eq.~\eqref{fequationnewmodel} allows us to reconstruct the $\alpha$-Attractors at high energies as a polynomial correction to the Starobinsky model, cf. Eq.~\eqref{polcorr}:
\begin{equation}\label{polcorrStaro}
	f(R) = R + \left[\frac{(1 - \beta)^2}{3M^2}\right]^\beta R^{\beta + 1} + \frac{(1 - \beta)^2}{6M^2}R^2\;.
\end{equation} 
This is just an approximation because the other sub-leading powers do not compensate, being indeed Eq.~\eqref{polcorr} not a solution of Eq.~\eqref{fequationbeta}.

If $n \ge 2$ in Eq.~\eqref{fequationnewmodel}, one can easily check that the equation can be balanced only if $n = 2$. This is expected because, as we have already commented earlier, in order to provide a scalar potential with a plateau at high energies, a $f(R)$ theory must go asymptotically as $R^2$.

\subsection{The $\alpha$-Attractors: T-models}

\begin{figure}[htbp]
	\centering
	\includegraphics[scale=0.5]{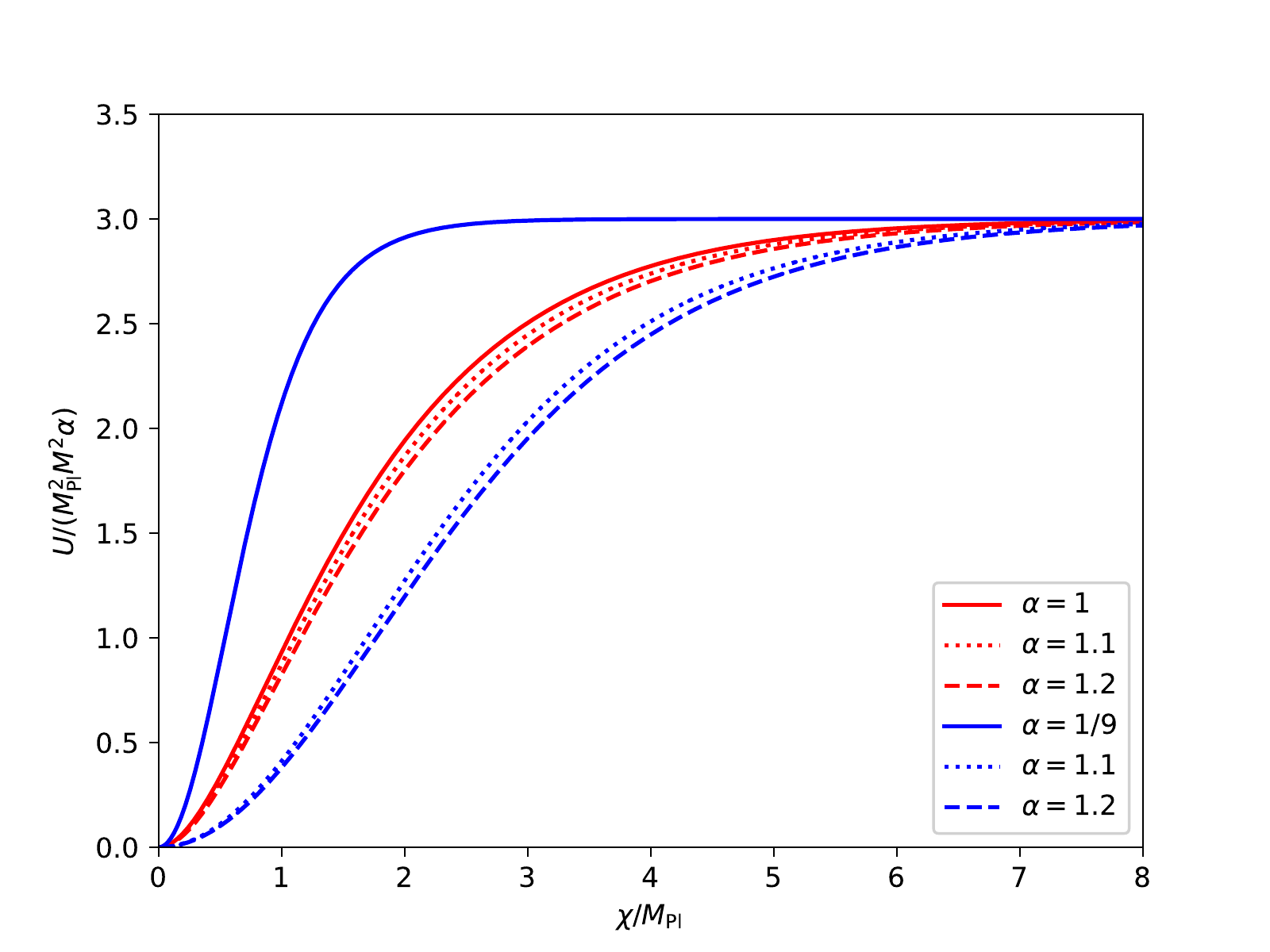}\includegraphics[scale=0.4]{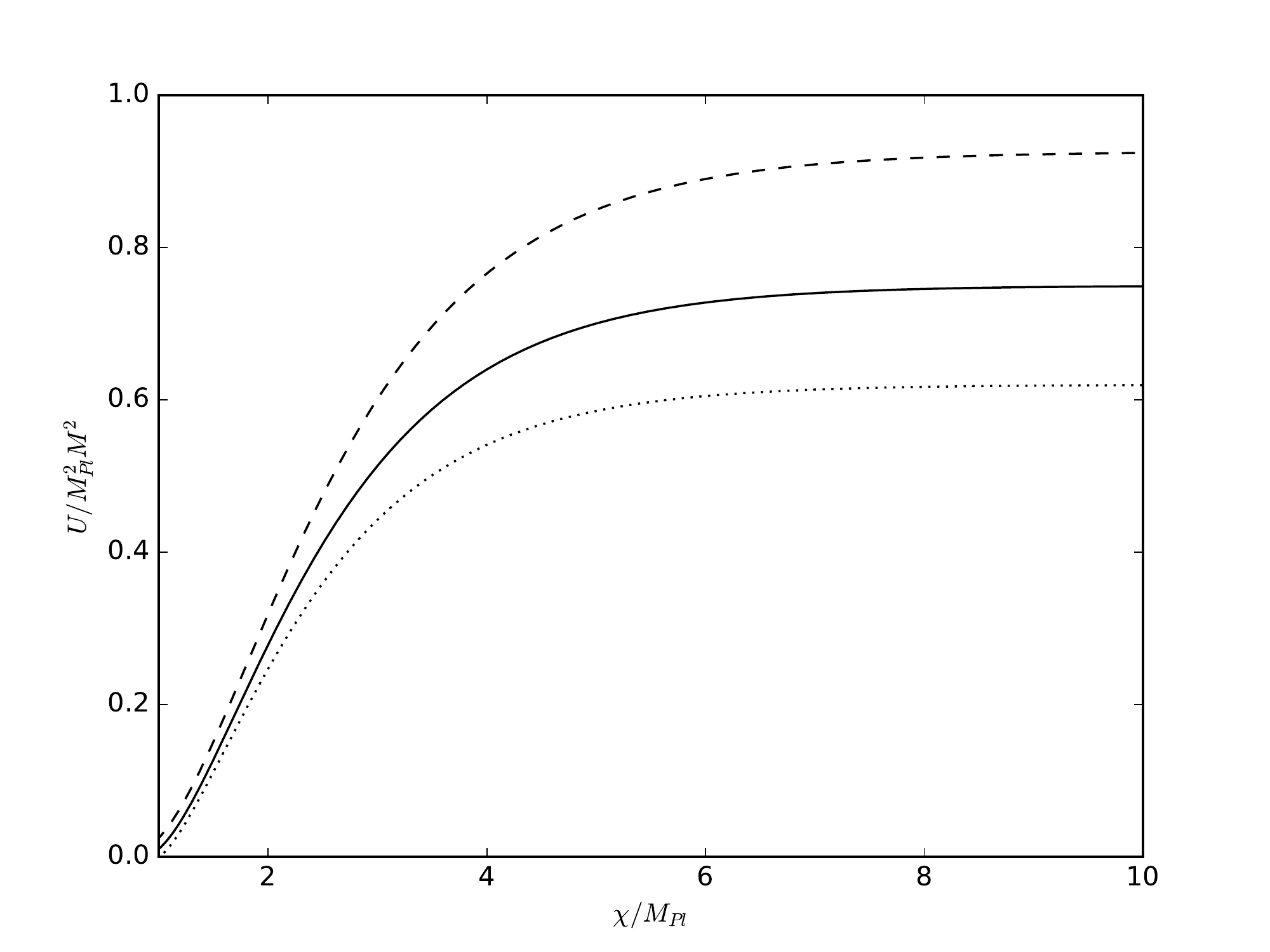}
	\caption{\textit{Left Panel.} Evolution of the E-Models potentials \eqref{alphattpotential} (red lines) and of the T-models potentials \eqref{tpot} (blue lines) for some values of $\alpha$. The choices $\alpha = 1$ and $\alpha = 1/9$ are interesting because they reproduce the Starobinsky model and the chaotic inflation model \cite{Goncharov:1983mw, Goncharov:1985yu}.\newline \textit{Right Panel.} Evolution of the normalised potential derived from Eq.~\eqref{polcorrStaronorm}. The solid line is drawn for the value $\beta = 1$, i.e. the Starobinsky model. The dashed and the dotted line are for $\beta = 0.1$ and $\beta = -0.1$, respectively.}
	\label{Fig:Uet}
\end{figure}

There exists another subclass of the $\alpha$-Attractors named T-Models. The potential characterising these models is the following:
\begin{eqnarray}\label{tpot}
 U_{T}(\chi) = 3M^{2}M_{\rm Pl}^2\alpha \tanh^{2}\left(\frac{\chi}{\sqrt{6\alpha}M_{\rm Pl}}\right)\;,
\end{eqnarray}
i.e. based on the plateau-like behavior of the hyperbolic tangent for large values of the field. Indeed, the above potential recovers the E-models potential, and therefore the Starobinsky one when $\alpha = 1$, only for large values of the field. In Fig.~\ref{Fig:Uet} we display the evolution of the E-models and T-models for some values of $\alpha$ including $\alpha = 1$, i.e. the Starobinsky model.

In the large $\alpha$ limit, by performing a Taylor-series expansion, it is easy to see that the T-models potential \eqref{tpot} coincides with the one of the chaotic inflation model with a quadratic potential \cite{Kallosh:2013hoa}, as for the E-models. 

As we did in the previous subsection, using the definitions given in Eqs.~\eqref{chidef} and \eqref{Upot} for the T-Model potential of Eq.~\eqref{tpot}, we can obtain a differential equation which allows to reconstruct a corresponding $f(R)$:
\begin{eqnarray}
f'^{2}\frac{\left[1 - f'^{(\beta - 1)}\right]^{2}}{\left[1 + f'^{(\beta - 1)}\right]^{2}} = \frac{(1 - \beta)^{2}}{6M^2}(Rf' - f)\;.
\end{eqnarray}
In the limit of large fields, i.e. $f^{'\beta - 1} \to 0$,\footnote{This limit holds true only if $f'$ is a growing function of $R$, i.e. $f'' > 0$, which is a condition that we are assuming.} the above equation becomes similar to Eq.~\eqref{fequationbeta}. Therefore, the same asymptotic analysis which led us to justify the ansatz given in Eq.~\eqref{polcorrStaro} applies.


\section{A power law extension of the Starobinsky model}\label{sec:extStaroalpha}

Motivated by the result of the previous section, here we address in detail the model given by Eq.~\eqref{polcorrStaro}, i.e. a power law extension of the Starobinsky model. In this case it is not possible to find a closed, analytic form for the potential of Eq.~\eqref{Upot} as function of the field $\chi$ of Eq.~\eqref{chidef}. In Refs.~\cite{Sebastiani:2013eqa, Sebastiani:2015kfa} the authors show that the following potential (written here in our notation):
\begin{equation}
	U(\chi) = M_{\rm Pl}^2\left(\alpha_1 - \gamma_1 e^{-n\sqrt{2/3}\chi/M_{\rm Pl}}\right)\;, 
\end{equation}
where $\alpha_1$ and $\gamma_1$ are parameter with dimension of a square mass, is able to reproduce asymptotically the following $f(R)$ model:
\begin{equation}
	f(R) \sim \frac{R^2}{4\gamma_1(2 - n)} + \frac{1}{2 - n}\left[\frac{1}{2\gamma_1(2 - n)}\right]^{1 - n}R^{2 - n}\;,
\end{equation}
via an analysis similar to the one performed in the previous section. From the above potential, it is possible to derive the following predictions on the scalar index and the tensor-to-scalar ratio:
\begin{equation}
	n_{\rm s} \simeq 1 - \frac{2}{N}\;, \qquad r \simeq \frac{12}{n^2N^2}\;,
\end{equation}
that is, a behaviour similar to the one predicted by the $\alpha$-Attractors, cf. Eq.~\eqref{alphaAttpred}.

Our analysis of the model \eqref{polcorrStaro} shall be purely numerical. The derivative of Eq.~\eqref{polcorrStaro}, is the following:
\begin{equation}\label{polcorrStarofp}
	f'(R) = 1 + (1 - \beta)^{2\beta}(1 + \beta)\left(\frac{R}{3M^2}\right)^{\beta} + (1 - \beta)^2\frac{R}{3M^2}\;.
\end{equation} 
This expression suggests the use of the dimensionless variable $R/3M^2$. Indeed, one can easily show that also the potential \eqref{Upot} is a function of $R/3M^2$, since:
\begin{equation}
	U = \frac{3M^2M_{\rm Pl}^2}{2(f')^2}\left(\frac{R}{3M^2}f' - \frac{f}{3M^2}\right)\;,
\end{equation}
and we have just shown that $f'$ is function of $R/3M^2$ and 
\begin{equation}\label{polcorrStaronorm}
	\frac{f}{3M^2} = \frac{R}{3M^2} + (1 - \beta)^{2\beta}\left(\frac{R}{3M^2}\right)^{\beta + 1} + \frac{(1 - \beta)^2}{2}\left(\frac{R}{3M^2}\right)^2\;,
\end{equation} 
is evidently function of $R/3M^2$. In Fig.~\ref{Fig:Uet} we display the evolution of the potential corresponding to Eq.~\eqref{polcorrStaro}.

\begin{figure}[htbp]
	\centering
	\includegraphics[scale=0.5]{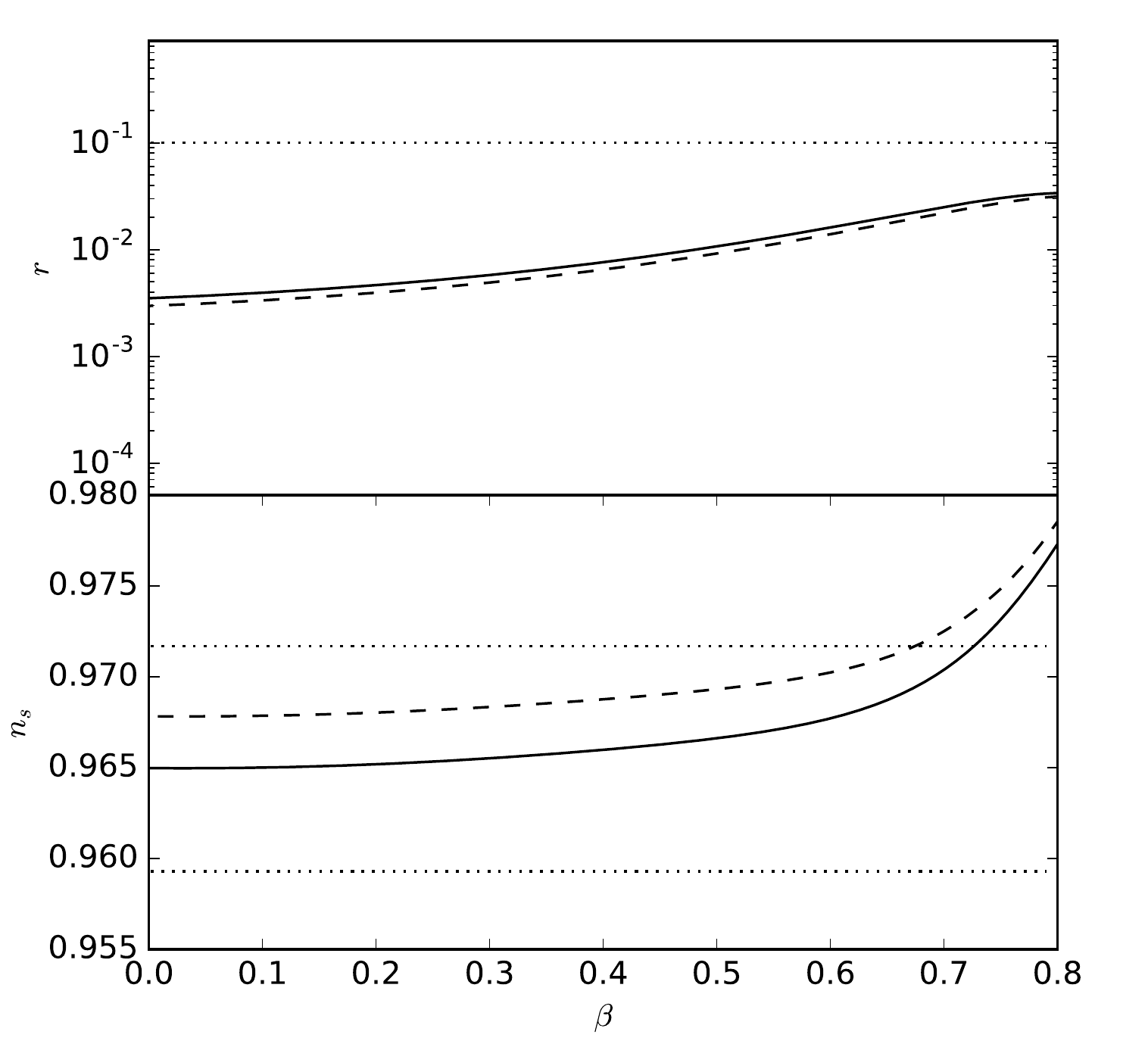}\includegraphics[scale=0.5]{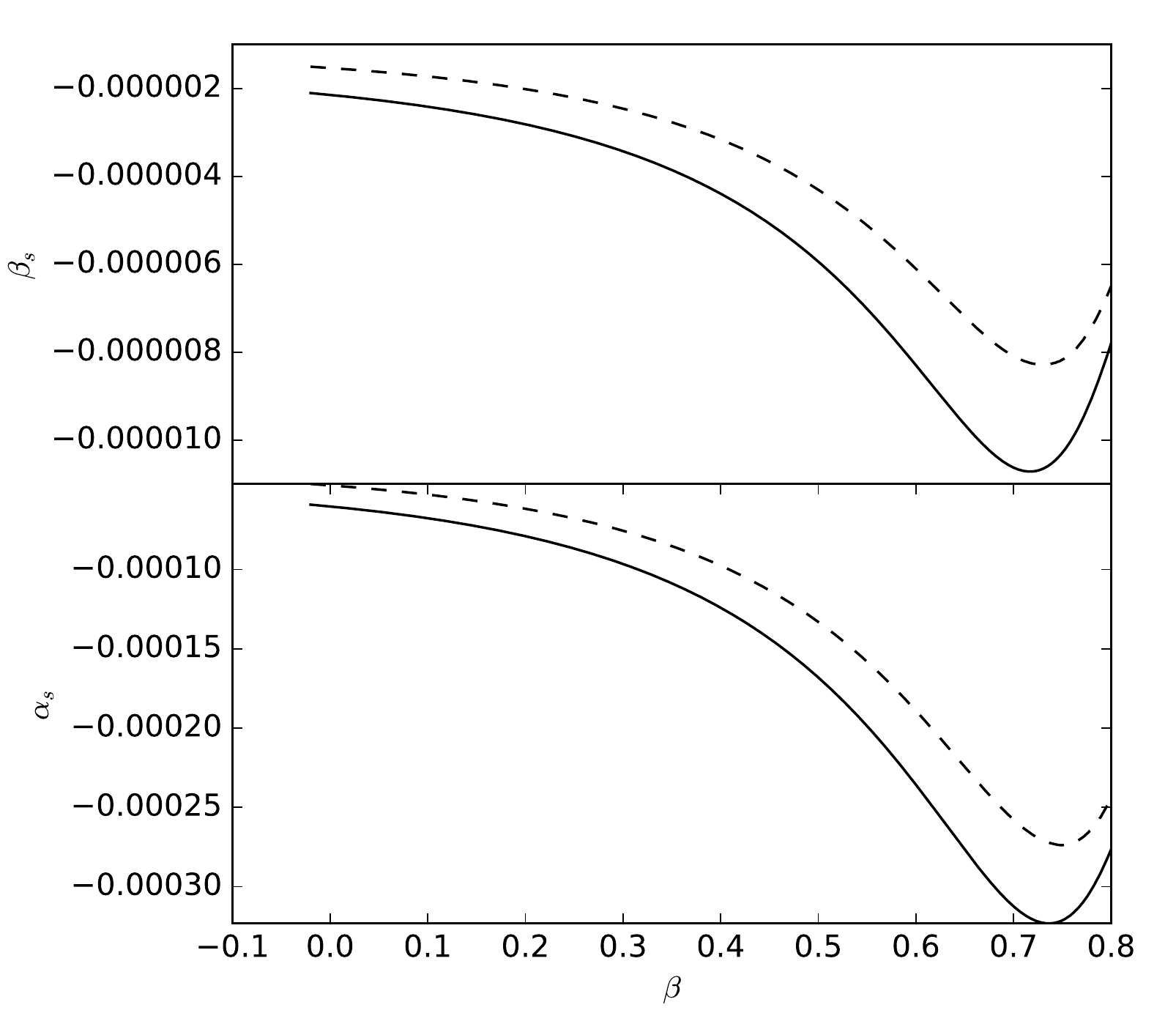}
	\caption{\textit{Left Panel.} Evolution of the inflationary parameters $n_{\rm s}$ and $r$ as functions of $\beta$ derived from Eq.~\eqref{polcorrStaronorm}. The solid line is drawn for $N = 55$ whereas the dashed line represents the case $N = 60$. The dotted lines represent the observational constraints at 95\% CL.\newline \textit{Right Panel.} Evolution of the runnings as functions of $\beta$ derived from Eq.~\eqref{polcorrStaronorm}. The solid line is drawn for $N = 55$ whereas the dashed line represents the case $N = 60$.}
	\label{Fig:nsrbetaplot}
\end{figure}
\begin{figure}[htbp]
	\centering
	\includegraphics[scale=0.6]{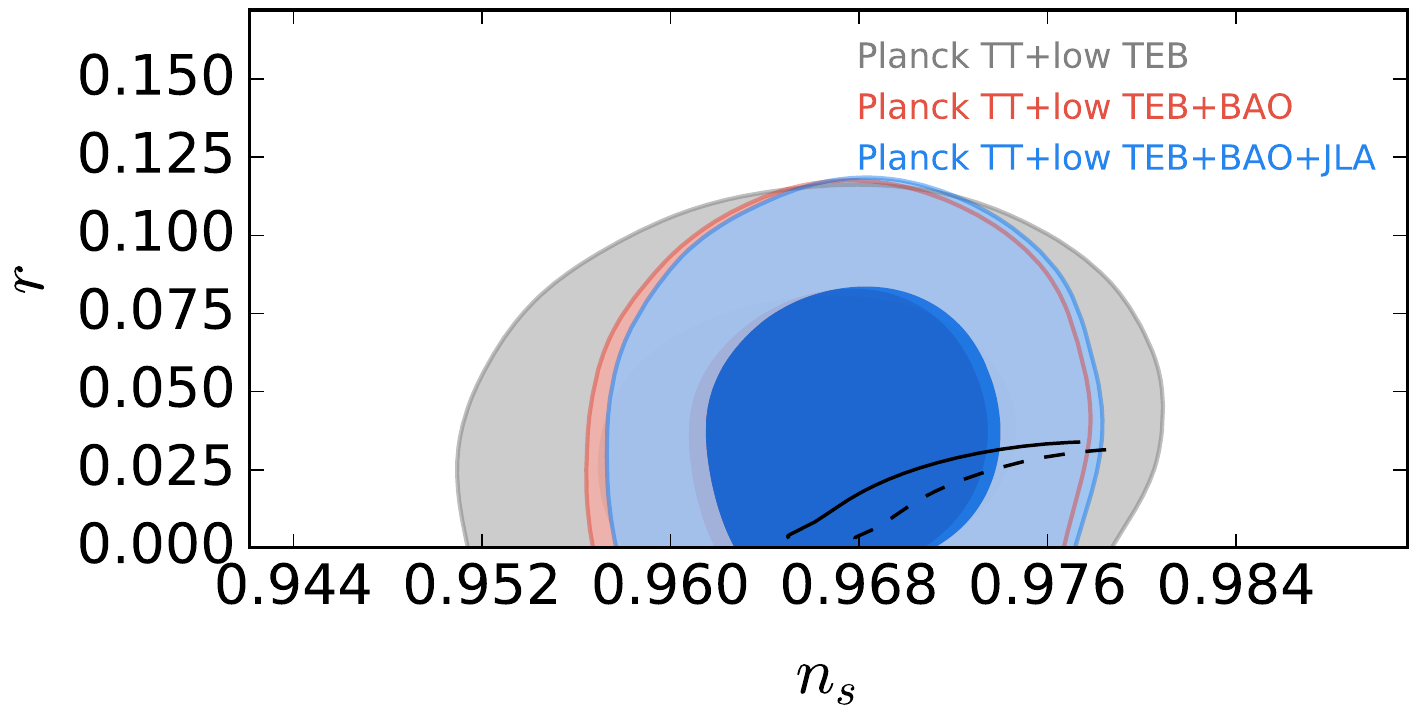}\\
	\caption{Evolution of $r$ \textit{vs} $n_{\rm s}$, varying $\beta$ in the interval $-0.02 < \beta < 0.8$ (corresponding to $0.96 < \alpha < 25$) and for $N = 55$ (solid line) and $N = 60$ (dashed line) with the marginalized $68\%$ and $95\%$ confidence level contours from Planck 2015 data \cite{Ade:2015lrj}.}
	\label{Fig:rnsbetaplot}
\end{figure}

The slow roll parameters can be defined as follows, given the functional form of the potential $U(\chi)$:
\begin{equation}
	\epsilon_U \equiv \frac{M_{Pl}^2}{2}\left(\frac{U_\chi}{U}\right)^2\;, \quad \eta_U \equiv M_{Pl}^2\frac{U_{\chi\chi}}{U}\;, \quad \xi_{U} \equiv M_{\rm Pl}^{4}\frac{U'U'''}{U^{2}}\;, \quad \sigma_{U} \equiv M_{\rm Pl}^{6}\frac{U'^{2}U^{(4)}}{U^{3}}\;,
\end{equation}
where $U^{(4)}$ represents the fourth derivative of the potential with respect to the field. From these quantities it is straightforward to compute the scalar index and the tensor-to-scalar ratio:
\begin{equation}
	n_{\rm s} \equiv 1 - 6\epsilon_U + 2\eta_U\;, \quad r \equiv 16\epsilon_U\;,
\end{equation}
and the running and the running of the running of the scalar spectral index:
\begin{eqnarray}
\alpha_{\rm s} &\equiv & \frac{dn_{\rm s}}{d\log k} = -2\xi_{U}^{2} + 16\eta_{U}\epsilon_{U} - 24\epsilon_{U}^{2}\;,\\
\beta_{\rm s} &\equiv & \frac{d\alpha_{\rm s}}{d\log k} = 2\sigma_{U}^{3} + 2\xi_{U}^{2}(\eta_{U} - 12\epsilon_{U}) - 32\epsilon_{U}(\eta_{U}^{2} - 6\eta_{U}\epsilon_{U} + 6\epsilon_{U}^{2})\;.
\end{eqnarray} 
For single field inflationary models, the above runnings are of the order of $\alpha_{\rm s}\sim 10^{-3}$ and $\beta_{\rm s}\sim 10^{-5}$ ~\cite{Garcia-Bellido:2014gna, Longden:2017iei}. Discriminating among $\alpha_{\rm s}$ and $\beta_{\rm s}$ of different models is still an experimental challenge, however it might be possible to use $\alpha_{\rm s}$ to this purpose in forthcoming Stage-4 CMB experiments, see e.g. Ref.~\cite{Munoz:2016owz}. In Fig.~\ref{Fig:nsrbetaplot} we display the evolution of $n_{\rm s}$ and $r$ as functions of $\beta$, showing a good agreement with the observational constraints for a range $0 < \beta \lesssim 0.8$. The latter value corresponds to a $R^{1.8}$ correction to the Starobinsky model. In Fig.~\ref{Fig:nsrbetaplot} we also plot the runnings as functions of $\beta$ and, finally, in Fig.~\ref{Fig:rnsbetaplot} we display the prediction of the model investigated in this section on the $r$ \textit{vs} $n_{\rm s}$ plane by varying $\beta$ in the interval $-0.02 < \beta < 0.8$ (corresponding to $0.96 < \alpha < 25$) and comparing this evolution with the contour regions allowed from the Planck data \cite{Ade:2015lrj}. As one can appreciate, the polynomial correction to $R^2$ allows larger values of $r$. In particular, it seems that diminishing the number of e-folds could allow to even larger values of $r$. 

\section{Study of the single power law $f(R)$ inflationary model}\label{Sec:singpowlaw}

In this section we address, for completeness, the case of a single power-law $f(R)$:
\begin{equation}\label{newmodel}
f(R) = R + \frac{R^{2 - \delta}}{(6M^2)^{1 - \delta}}\;,
\end{equation}
and compute its predictions on the inflationary parameters. This kind of model has already been investigated in Ref.~\cite{Motohashi:2014tra}, in order to test the robustness of the Starobinsky model and assess how much precise data have to be in order to detect deviations from the case $\delta = 0$. We present a similar analysis in next section, providing analytic results for the inflationary parameters $n_{\rm s}$ and $r$ as functions of the number of e-folds $N$ and the new parameter $\delta$.

\begin{figure}[htbp]
\centering
	\includegraphics[scale=0.5]{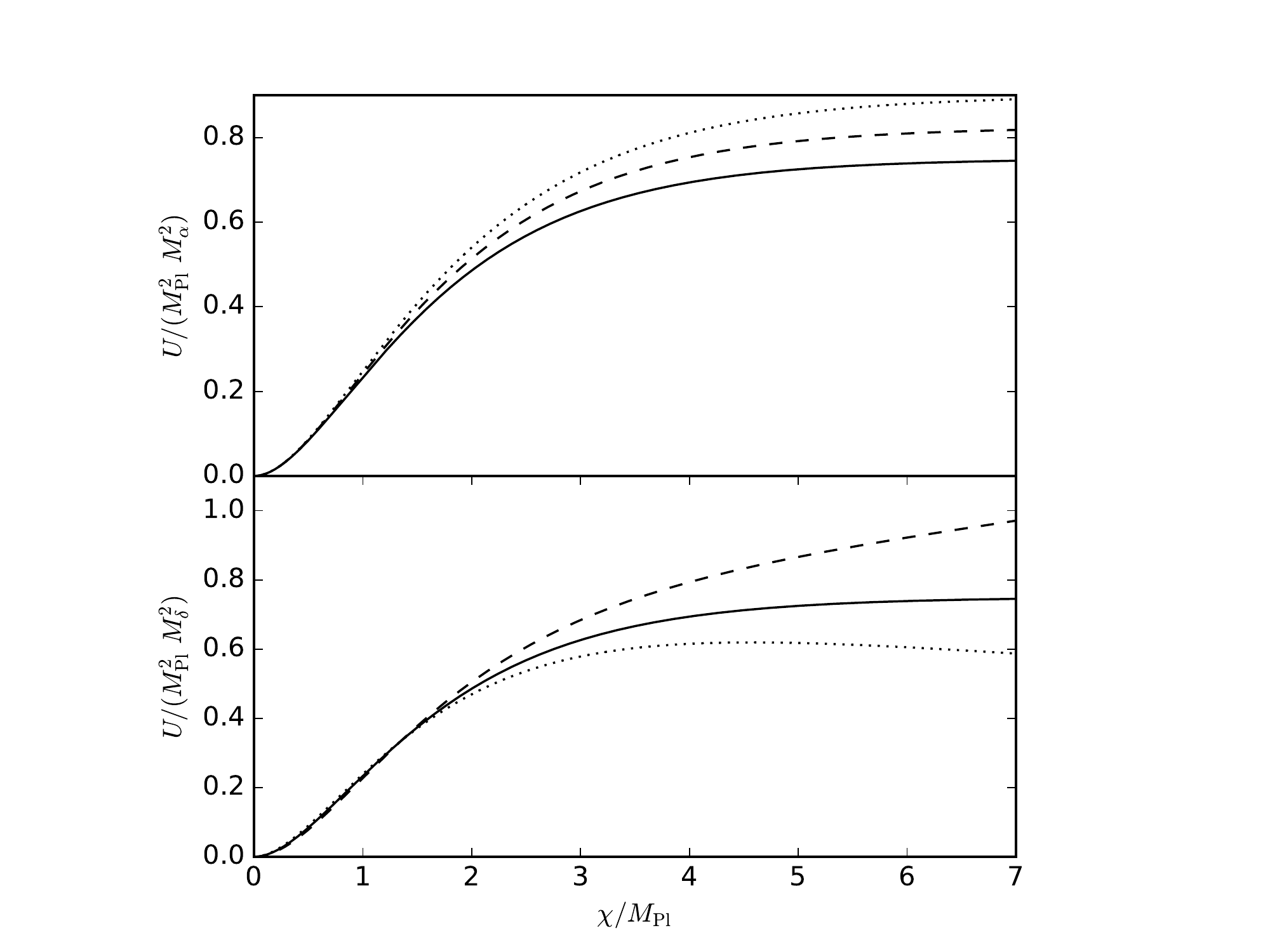}
	\caption{Upper panel.  Evolution of the potential \eqref{alphattpotential} for $\alpha = 1.1$ (dashed line) and $\alpha = 1.2$ (dotted line) compared with the Starobinsky case (solid line). Lower panel. Evolution of the potential \eqref{deltapot} for $\delta = 0.05$ (upper dashed line) and $\delta = -0.05$ (lower dotted line) compared with the Starobinsky case (solid line).
		}
	\label{Fig:Uplots}
\end{figure}

Starting from Eq.~\eqref{newmodel}, we can determine the following potential for the scalar field $\chi$:
\begin{equation}\label{deltapot}
U(\chi) = \frac{3M_{\rm Pl}^2M^2(1 - \delta)}{(2-\delta)^{\frac{2 - \delta}{1 - \delta}}}e^{-2\sqrt{\frac{2}{3}}\frac{\chi}{M_{\rm Pl}}}\left(e^{\sqrt{\frac{2}{3}}\frac{\chi}{M_{\rm Pl}}} - 1\right)^{\frac{2 - \delta}{1 - \delta}}\;.
\end{equation}
It is straightforward to see that for $\delta = 0$ we recover the Starobinsky model given in Eq.~\eqref{Staropotential}. For large fields, the above potential behaves as:
\begin{equation}
U(\chi) \sim \frac{3M_{\rm Pl}^2M^2(1 - \delta)}{(2-\delta)^{\frac{2 - \delta}{1 - \delta}}}e^{\sqrt{\frac{2}{3}}\frac{\delta}{1 - \delta}\frac{\chi}{M_{\rm Pl}}}\;, \qquad (\chi \gg M_{\rm Pl})\;,
\end{equation}
where again, for $\delta = 0$, we recover the plateau typical of the Starobinsky model. 
In Fig.~\ref{Fig:Uplots} we show the behaviour of $U(\chi)$ of the model \eqref{deltapot} for different values of $\delta$ and of \eqref{alphattpotential} for different values of $\alpha$, compared with the Starobinsky model.

As expected, from Fig.~\ref{Fig:Uplots} one can see that the $\alpha$-Attractors potential seems to preserve the plateau at high energies, whereas the $\delta$-model displays an important change in the steepness of the potential. This statement can be made more quantitative by calculating the slow roll parameters from Eq.~\eqref{deltapot}. One has for $\epsilon_U$:
\begin{eqnarray}
\epsilon_U \equiv \frac{M_{\rm Pl}^2}{2}\left(\frac{U_\chi}{U}\right)^2 = \frac{1}{3}\left(\frac{1}{1 - \delta}\right)^2\left[\frac{\delta + 2(1 - \delta)e^{-\sqrt{\frac{2}{3}}\frac{\chi}{M_{\rm Pl}}}}{1 - e^{-\sqrt{\frac{2}{3}}\frac{\chi}{M_{\rm Pl}}}}\right]^2\;,  
\end{eqnarray}
and $\eta_U$:
\begin{equation}
	\eta_U \equiv M_{\rm Pl}^2\frac{U_{\chi\chi}}{U} = \frac{2}{3(1 - \delta)^2}\frac{\delta^2 - (1 - \delta)(2 - 5\delta)e^{-\sqrt{\frac{2}{3}}\frac{\chi}{M_{\rm Pl}}} +  4(1 - \delta)^2e^{-2\sqrt{\frac{2}{3}}\frac{\chi}{M_{\rm Pl}}}}{\left(1 - e^{-\sqrt{\frac{2}{3}}\frac{\chi}{M_{\rm Pl}}}\right)^2}
\end{equation}
%
For large fields $(\chi \gg M_{\rm Pl})$ the slow roll parameters can be approximated as follows:
\begin{equation}
\epsilon_U \sim \frac{1}{3}\left(\frac{1}{1 - \delta}\right)^2\left[\delta + (2 - \delta)e^{-\sqrt{\frac{2}{3}}\frac{\chi}{M_{\rm Pl}}}\right]^2\;,
\end{equation}
and
\begin{equation}
\eta_U \sim \frac{2}{3(1 - \delta)^2}\left[\delta^2 - (2 - 7\delta + 3\delta^2)e^{-\sqrt{\frac{2}{3}}\frac{\chi}{M_{\rm Pl}}}\right]\;.
\end{equation}
Note that 
\begin{equation}
\epsilon_U \to \frac{1}{3}\left(\frac{\delta}{1 - \delta}\right)^2\;, \qquad \chi \to \infty\;,
\end{equation}
i.e. there is no plateau if $\delta > 0$, as we expected from observing Fig.~\ref{Fig:Uplots} and from the analysis performed earlier in Sec.~\ref{sec:Staromodel}. Moreover, since $\epsilon_U \ll 1$ in order to have an inflationary phase, then $\delta \ll 1$. 

\begin{figure}[htbp]
	\centering
	\includegraphics[scale=0.5]{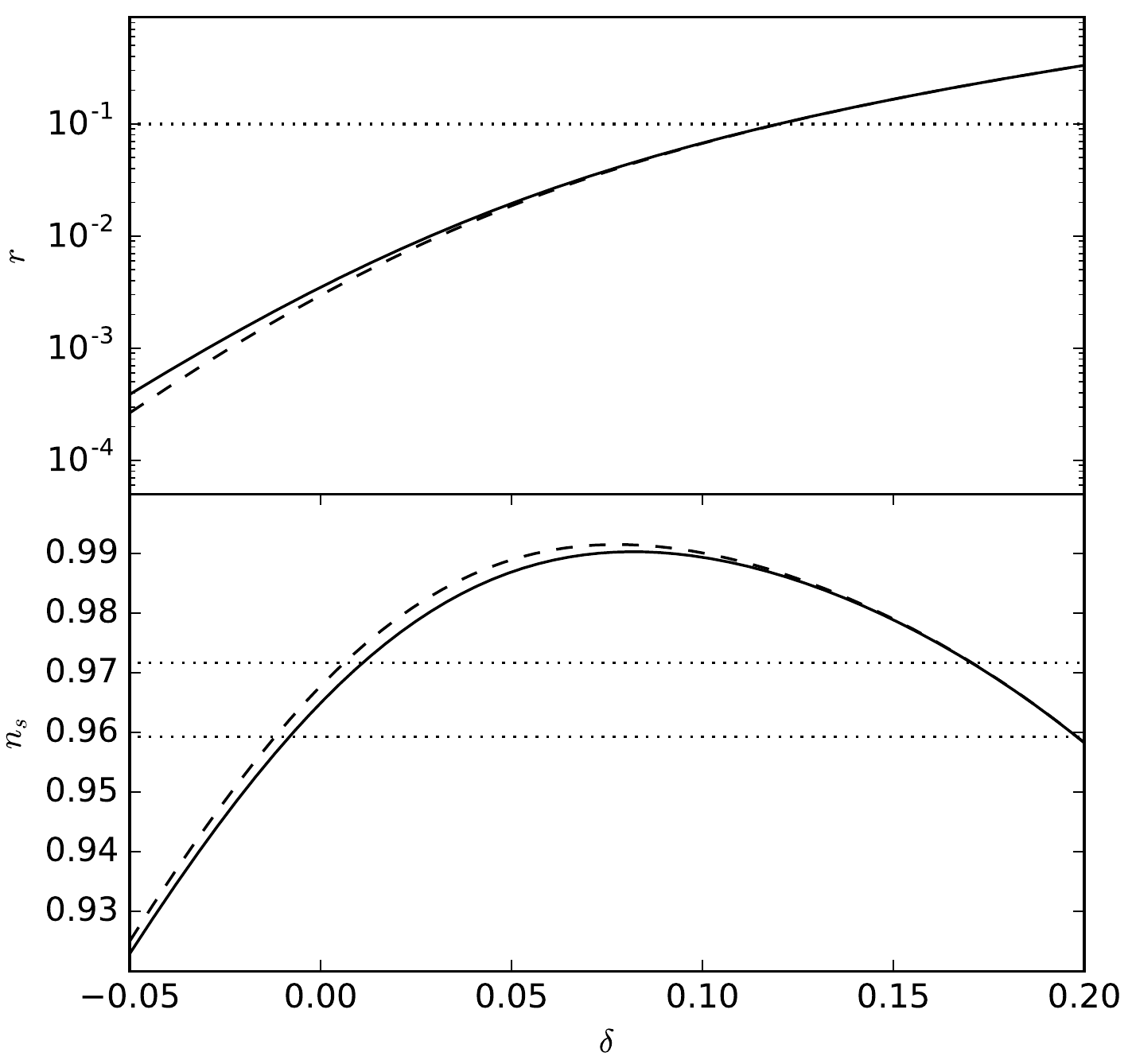}\includegraphics[scale=0.5]{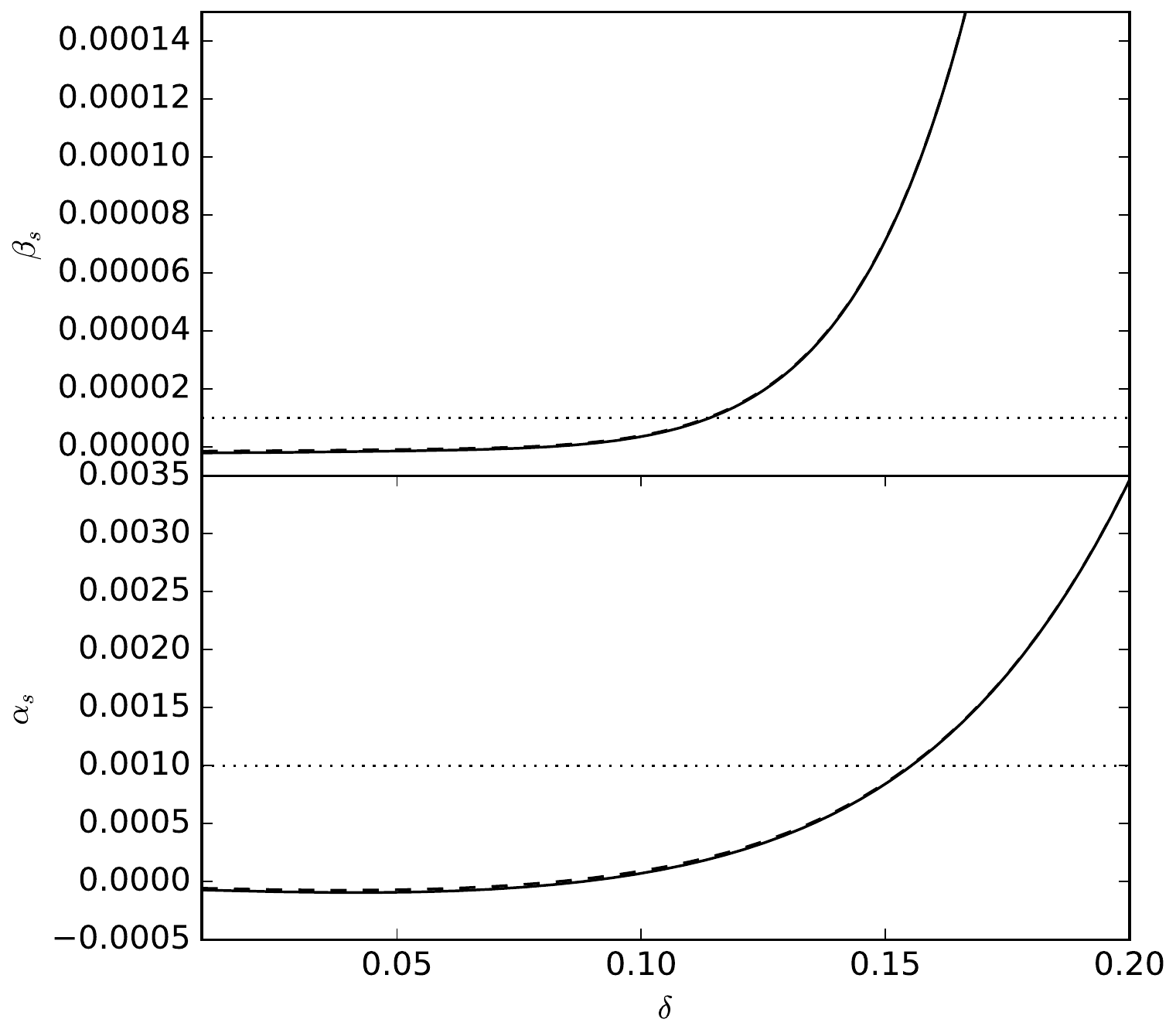}
	\caption{Evolution of $r$, $n_{\rm s}$,  $\beta_{\rm s}$ and $\alpha_{\rm s}$ as functions of $\delta$ for $N = 55$ (solid line) and $N = 60$ (dashed line). The horizontal dotted line is the upper limit on $r$, at 95\% confidence level, obtained by the Planck collaboration \cite{Ade:2015lrj}. The horizontal dotted lines enclose the 68\% confidence level of the values of $n_{\rm s}$ measured by the Planck collaboration \cite{Ade:2015lrj}
		 .}
	\label{Fig:nsrdeltaplots}
\end{figure}

The value of the field at which inflation ends is given by $\epsilon_U(\chi_f) \approx 1$, which in our case translates to:
\begin{equation}
\epsilon_U(\chi_f) \approx 1 \quad \Rightarrow \quad e^{-\sqrt{\frac{2}{3}}\frac{\chi_f}{M_{\rm Pl}}} \approx \frac{\sqrt{3} - \delta(1 + \sqrt{3})}{(2 + \sqrt{3})(1 - \delta)} \simeq \frac{\sqrt{3} - \delta}{2 + \sqrt{3}}\;.
\end{equation}
With this we can calculate the number of e-folds as follows:
\begin{equation}
N = \frac{1}{\sqrt{2}M_{\rm Pl}}\int_{\chi_f}^{\chi_i}\frac{d\chi}{\sqrt{\epsilon_U}}\;.
\end{equation}
Let's stay at the lowest possible order in both $\delta$ and $e^{-\sqrt{\frac{2}{3}}\frac{\chi}{M_{\rm Pl}}}$ and their combinations. In this case, we can approximate the slow roll parameters and the number of e-folds as follows:
\begin{eqnarray}
 \epsilon_U \sim \frac{1}{3}\left(\delta + 2e^{-\sqrt{\frac{2}{3}}\frac{\chi}{M_{\rm Pl}}}\right)^2\;, \quad \eta_U \sim
	 -\frac{4}{3}e^{-\sqrt{\frac{2}{3}}\frac{\chi}{M_{\rm Pl}}}\;, \quad \frac{1}{N} \sim \frac{4}{3}e^{-\sqrt{\frac{2}{3}}\frac{\chi}{M_{\rm Pl}}} + \frac{\delta}{3}\;.
\end{eqnarray}
The scalar index and the tensor-to-scalar ratio are therefore written as:
\begin{eqnarray}
	&&n_{\rm s} \equiv 1 - 6\epsilon_U + 2\eta_U \sim 1 - \frac{2}{N}\left(1 - \frac{N\delta}{3}\right)\;,\\
	&&r \equiv 16\epsilon_U \sim \frac{12}{N^2}\left(1 + \frac{N\delta}{3}\right)^2\;.
\end{eqnarray}
As we can see, differently from the $\alpha$-Attractors case, the $\delta$ correction interferes also with the scalar spectral index, through the combination $N\delta$. In Fig.~\ref{Fig:nsrdeltaplots} we display the numerical results for the evolution of $n_{\rm s}$ and $r$ as functions of $\delta$, by fixing the number of e-folds $N = 55$ and $N = 60$. For completeness, in the same figure we also calculate the nonvanishing runnings $\alpha_{\rm s}=dn_{\rm s}/d\log k$ and $\beta_{\rm s}=d\alpha_{\rm s}/d\log k$.

\begin{figure}[htbp]
	\centering
	\includegraphics[scale=0.6]{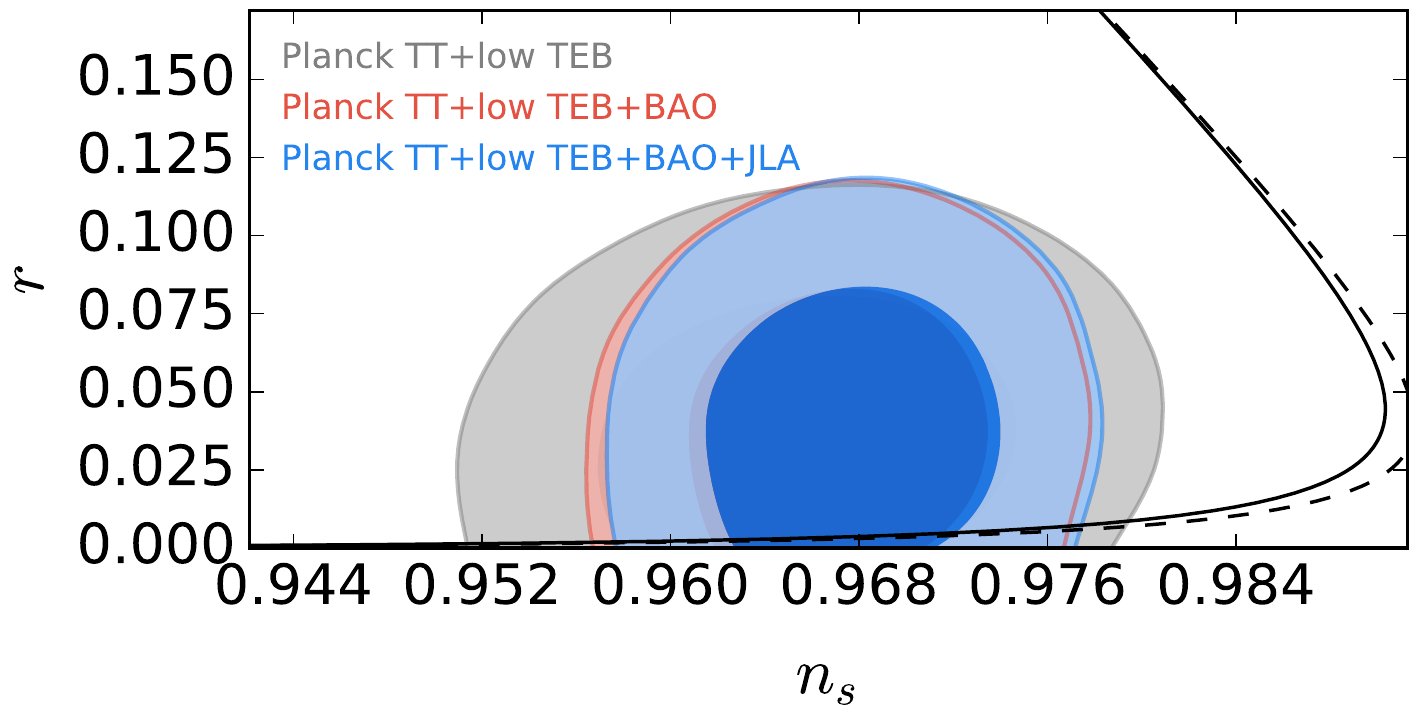}\\
	\caption{Evolution of $r$ \textit{vs} $n_{\rm s}$, varying $\delta$ and for $N = 55$ (solid line) and $N = 60$ (dashed line) with the marginalized $68\%$ and $95\%$ confidence level contours from Planck 2015 data.}
	\label{Fig:rnsplot}
\end{figure}

In Fig.~\ref{Fig:rnsplot} we display the $\delta$ model in the $r$ \textit{vs} $n_{\rm s}$ plane, again choosing $N = 55$ and $N = 60$.  When $N=55$, the variation of the $\delta$ parameter is $-0.004<\delta<0.011$ for $1\sigma$, and $-0.011<\delta<0.022$ for $2\sigma$. When $N=60$, it is $-0.008<\delta<0.006$ for $1\sigma$, and $-0.015<\delta<0.016$ for $2\sigma$. This figure is very similar to the first panel of Fig. 4 of Ref.~\cite{Motohashi:2014tra}.

Differently from the $\alpha$-Attractors case, the $\delta$ model move in the ``wrong" direction in the $r$ \textit{vs} $n_{\rm s}$, i.e. they move mostly horizontally.


\section{Discussion and conclusions}\label{sec:concl}

Motivated by the good agreement between the cosmological parameters $n_{\rm s}$ and $r$ provided by the $\alpha$-Attractors theory and recent observations, we derive a differential equation for a $f(R)$ theory from the scalar potentials of $\alpha$-Attractors, cf. Eq.~\eqref{fequationbeta}, in order to find a $f(R)$ theory compatible with this class of models. Since this differential equation cannot be solved analytically for any choice $\alpha$, we make an asymptotic analysis at high energies, where Inflation is expected to take place, and show that the ansatz \eqref{polcorrStaro} represents a viable solution to the differential equation \eqref{fequationbeta} at the leading and sub-leading order in the limit $R \to \infty$.

Then, we investigate the predictions on the inflationary parameters provided by the power law extension of the Starobinsky model and show that they are in very good agreement with observation, allowing for a greater production of gravitational waves, i.e. a larger $r$, than in the Starobinsky model.

We also made a detailed analysis of the model $f(R) = \gamma R^{2 - \delta}$ and found a mostly horizontal movement of the predictions of the model in the $r$--$n_{\rm s}$ plane, indicating that the missing of the $R^2$ term in the action alters drastically the behaviour of the inflationary evolution.


The reconstruction here performed of the $\alpha$-Attractors models is confined to a $f(R)$ model. A future investigation in this sense could start from a more general extension of GR, such as the Horndeski theory \cite{Horndeski:1974wa}.

\begin{acknowledgments}

This work has been supported by CNPq (Brazil), CAPES (Brazil) and Fapes (Brazil). We thank Rodrigo vom Marttens and Jhonny Andres Agudelo Ruiz for useful discussions. We are also indebted to professors D. Wands, A. A. Starobinsky and J. D. Barrow for enlightening discussions and precious suggestions. Finally, we would like to thank the users of Mathematica Stack Exchange who helped us in dealing with the numerical analysis of Eq.~\eqref{fequationbeta}. 
\end{acknowledgments}

\appendix

\section{Detailed investigation of the differential equation \eqref{fequationbeta}}\label{App}

In this appendix we investigate in greater detail Eq.~\eqref{fequationbeta}, which we report here:
\begin{equation}\label{fequationbeta2}
Rf' - f = \frac{3M^2}{2(1-\beta)^{2}}\left(f' - f^{'\beta}\right)^2\;.
\end{equation}
First of all, we define $f(R) \equiv R + F(R)$, i.e. we postulate the presence of the Einstein-Hilbert term and focus our attention just on its correction. Equation \eqref{fequationbeta2} thus becomes:
\begin{equation}
	RF' - F = \frac{3M^2}{2(1-\beta)^{2}}(1 + F')^2\left[1 - (1 + F')^{\beta - 1}\right]^2\;.
\end{equation} 
Next, we normalise $R$ and $F$ to $3M^2/2$ and define $\gamma \equiv \beta - 1 = -1/\sqrt{\alpha} < 0$, thus obtaining:\footnote{We discussed this equation in \url{https://mathematica.stackexchange.com/questions/152812/numerical-solution-to-a-nonlinear-ordinary-differential-equation}, obtaining very useful help.}
\begin{equation}\label{Fgammaeq}
	RF' - F = \frac{(1 + F')^2}{\gamma^2}\left[1 - (1 + F')^{\gamma}\right]^2\;.
\end{equation} 
Deriving this equation with respect to $R$, it is not difficult to obtain:
\begin{equation}
	\frac{2\left(F' + 1\right)\left[\left(F' + 1\right)^{\gamma} - 1\right]\left[(\gamma + 1)\left(F' + 1\right)^{\gamma} - 1\right]}{\gamma^2}F'' = R F''\;.
\end{equation}
Now, this equation is satisfied if $F'' = 0$ but then it would give a linear solution for $F(R)$ in which we are not interested since we have already stipulated that $f(R) \equiv R + F(R)$. Therefore, we assume that $F'' > 0$, and the above equation gives us:
\begin{equation}\label{Fpconstr}
	R = \frac{2\left(F' + 1\right)\left[\left(F' + 1\right)^{\gamma} - 1\right]\left[(\gamma + 1)\left(F' + 1\right)^{\gamma} - 1\right]}{\gamma^2}\;.
\end{equation}
For $\gamma = -1$, one can easily recover the Starobinsky case, since:
\begin{equation}\label{StarobehaviourF}
	R = 2F' \quad \Rightarrow \quad F = \frac{R^2}{4}\;,
\end{equation}  
and restoring the normalisation by $3M^2/2$ one obtains Eq.~\eqref{Staromodel}.

Equation \eqref{Fpconstr} gives us a constraint on the expression of $F'(R)$ which we must take into account when solving \eqref{Fgammaeq} in order to select the correct initial condition and exclude the linear solution. Note that $F' = 0$ implies from Eq.~\eqref{Fpconstr} that $R = 0$ and then, from Eq.~\eqref{Fgammaeq}, that $F = 0$. We have thus that $F(0) = F'(0) = 0$, as desired.

\begin{figure}[htbp]
	\centering
	\includegraphics[scale=0.5]{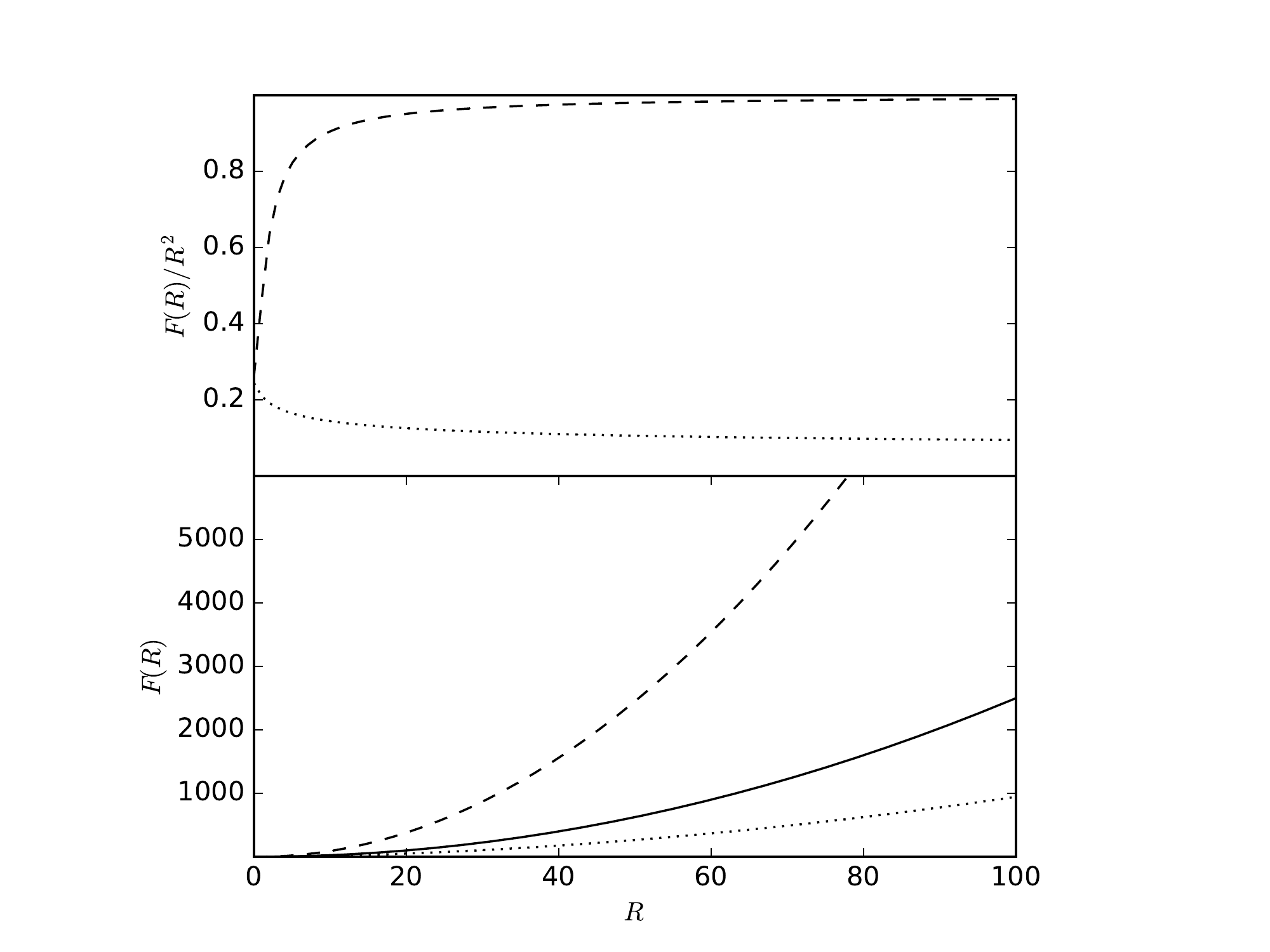}
	\caption{\textit{Upper Panel}. Evolution of the numerically reconstructed $F(R)$ normalised to $R^2$ for the cases $\gamma = -2$ ($\alpha = 1/4$, dashed line) and $\gamma = -1/2$ ($\alpha = 4$, dotted line). This plot shows an asymptotic quadratic behaviour.\newline \textit{Lower Panel}. Reconstructed $F(R)$ for the cases $\gamma = -2$ ($\alpha = 1/4$, dashed line), the Starobinsky model $\gamma = -1$ ($\alpha = 1$, solid line) and $\gamma = -1/2$ ($\alpha = 4$, dotted line). }
	\label{Fig:NumsolalphaAtt}
\end{figure}

The strategy in order to numerically obtain a solution $F(R)$ is the following. We choose an initial, small but non vanishing $R_i \equiv \epsilon$; determine $F'(\epsilon)$ from Eq.~\eqref{Fpconstr}; determine the initial condition $F(\epsilon)$ from Eq.~\eqref{Fgammaeq}; solve either Eq.~\eqref{Fgammaeq} or Eq.~\eqref{Fpconstr}. A simpler way is to use the fact that $F'(R) \to 0$ for $R \to 0$. Then, for a sufficiently small $\epsilon$, keeping the lowest order in $F'(\epsilon)$ in Eqs.~\eqref{Fpconstr} and \eqref{Fgammaeq}, we get:
\begin{equation}
	\epsilon \sim 2F'(\epsilon)\;, \qquad F(\epsilon) = \frac{\epsilon^2}{4}\;,
\end{equation} 
which is indeed the Starobinsky case found in Eq.~\eqref{StarobehaviourF}, revealing thereby that the Starobinsky model is a low curvature approximation of the $\alpha$-Attractors. This can be also seen by taking the low $\chi$ limit of Eq.~\eqref{alphattpotential}. We can also prove that the Starobinsky model is a high curvature approximation of the $\alpha$-Attractors, since $F'' > 0$ and $\gamma < 0$. Then, for $R \to \infty$ we get from Eq.~\eqref{Fpconstr}:
\begin{equation}
	R \sim \frac{2F'}{\gamma^2}\;, \quad \mbox{ for } R\to\infty\;, \quad \Rightarrow \quad F \sim \frac{\gamma^2}{4}R^2\;, \quad \mbox{ for } R \to \infty\;.
\end{equation}
Being $\gamma = \beta - 1$, restoring the $3M^2/2$ normalisation, one easily recover our approximation \eqref{polcorrStaro}. Indeed, the plots in the upper panel of Fig.~\ref{Fig:NumsolalphaAtt} suggest that the Starobinsky model is also a high curvature approximation of the $\alpha$-Attractors (of course, with different energy scales depending on the parameter $\alpha$). This fact is somehow expected, since both the theories are characterised by a plateau potential at high energies.

\begin{figure}[htbp]
	\centering
	\includegraphics[scale=0.5]{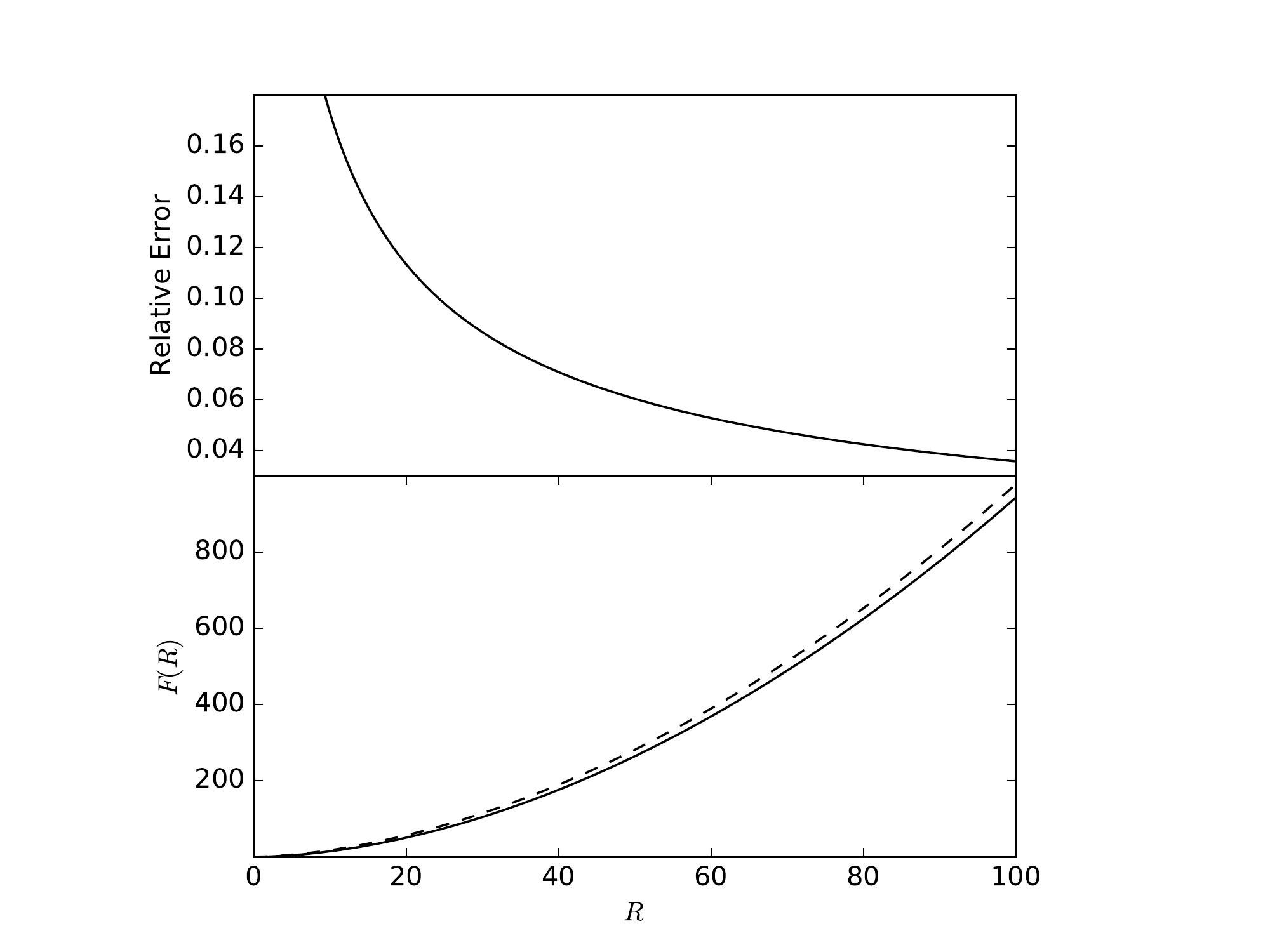}
		\caption{ \textit{Upper Panel.} Relative error, showing the goodness of our approximation Eq.~\eqref{polcorrStaro}.\newline \textit{Lower Panel}. Comparison between the numerical reconstructed $F(R)$ using the techniques of this section and the approximated one in Eq.~\eqref{polcorrStaro}, for $\gamma = -1/2$ (corresponding to $\beta = 1/2$ and $\alpha = 4$).}
	\label{Fig:betahalfcomp}
\end{figure}

In Fig.~\ref{Fig:betahalfcomp} we also display the goodness of our approximation \eqref{polcorrStaro} in the case $\alpha = 4$.



\bibliographystyle{unsrturl}
\bibliography{Staroalphaatt}

\end{document}